\documentclass[]{aastex631}%linenumbers
\usepackage{CJKutf8}
\usepackage{amsmath}
\usepackage{bm}

\newcommand{\tf}{t_{\rm frozen}}

\newcommand{\rj}{ R_{\rm 0}}
\newcommand{\rhoj}{\rho_{0}}

\received{ }
\revised{ }
\accepted{ }
\submitjournal{ }

\shorttitle{UHECR acceleration in kpc-scale jets}
\shortauthors{Wang et al.}

\begin{document}
\begin{CJK*}{UTF8}{gkai}

\title{Acceleration of ultra-high-energy cosmic rays in the kiloparsec-scale jets of nearby radio galaxies}% Force line breaks with 

\correspondingauthor{Jieshuang Wang}
\email{jswang@mpi-hd.mpg.de, ~jieshuang.wang@ipp.mpg.de}

\author[0000-0002-2662-6912]{Jie-Shuang Wang (王界双)}

\affiliation{Max-Planck-Institut f\"ur Kernphysik, Saupfercheckweg 1, D-69117 Heidelberg, Germany} %\textbackslash\textbackslash}%
\affiliation{Max Planck Institute for Plasma Physics, Boltzmannstra{\ss}e 2, D-85748 Garching, Germany}
\author[0000-0002-3778-1432]{Brian Reville}
\affiliation{Max-Planck-Institut f\"ur Kernphysik, Saupfercheckweg 1, D-69117 Heidelberg, Germany}
\author[0000-0003-1334-2993]{Frank M. Rieger}
\affiliation{Max Planck Institute for Plasma Physics, Boltzmannstra{\ss}e 2, D-85748 Garching, Germany}
\affiliation{Institute for Theoretical Physics, Heidelberg University, Philosophenweg 12, D-69120 Heidelberg, Germany}
\affiliation{Max-Planck-Institut f\"ur Kernphysik, Saupfercheckweg 1, D-69117 Heidelberg, Germany}
\author[0000-0003-1157-3915]{Felix A. Aharonian}
\affiliation{Dublin Institute for Advanced Studies, 31 Fitzwilliam Place, Dublin 2, Ireland}
\affiliation{Max-Planck-Institut f\"ur Kernphysik, Saupfercheckweg 1, D-69117 Heidelberg, Germany}
\affiliation{Yerevan State University, Alek Manukyan St 1, Yerevan 0025, Armenia}

\begin{abstract}
Radio galaxies have long been considered as potential sources of ultra-high-energy cosmic rays (UHECRs). Recent 
analyses of the UHECR spectrum, composition, and arrival directions indicate that the nearest radio galaxy, Centaurus A, 
could be linked to the reported dipole anisotropy, though the mechanism underlying the acceleration remains elusive. 
In this Letter, we explore UHECR acceleration in the kiloparsec-scale jets of radio galaxies, exemplified by Centaurus A. 
Using high-resolution relativistic magneto-hydrodynamic and test-particle simulations without sub-grid physics, we 
investigate the acceleration of the highest-energy particles in the turbulent sheath of a fast-moving jet. Our 
findings demonstrate that acceleration close to the maximum theoretical expectation is possible. When extrapolated 
to nearby radio galaxies, our results suggest that the kiloparsec-scale jets of Centaurus A could account for the dipole 
anisotropy in UHECRs, while more potent Fanaroff-Riley type II radio galaxies may account for the observed UHECR 
spectrum with a rigidity cutoff at a few Exavolts.
\end{abstract}

%\keywords{ }

\section{Introduction}
The origin of ultra-high-energy cosmic rays (UHECRs, $E\gtrsim1~{\rm EeV}=10^{18}$~eV) remains an unsolved
problem. Observationally, significant advances have been made recently, thanks to data collected by the 
Pierre Auger Observatory (PAO) and the Telescope Array experiment.
In particular, the extragalactic origin of UHECRs is now corroborated by the detection of a dipole anisotropy 
in the arrival directions of UHECRs \citep{PierreAugerCollaboration2017Sci,Aab2018ApJ}.
UHECR measurements \citep{Aab2020PhRvD,Aab2020PhRvL,Tsunesada:2021qO,Abbasi2023APh} have revealed several 
features in the all-particle spectrum, including the \textit{ankle} at $\approx 5$~EeV, an \textit{instep}
at $\approx13$~EeV, and a cut-off at $\approx50$~EeV, with individual events detected up to $\approx 300$ 
EeV \citep{Bird1995,Abbasi2023S}.
Combined analyses of the spectrum and composition data do favour a scenario in which the UHECR flux has a 
mixed composition, changing from protons to heavier elements with a hard spectrum and cutoff at a rigidity ($\equiv E/Ze$) 
of a few~EV \citep[e.g.,][]{Aab:JCAP:2017,Aab2020PhRvL,AbdulHalim2023JCAP}, where $Ze$ is the charge of a given atom.
When arrival directions are taken into account, the findings suggest that the nearby active galaxy 
Centaurus A, hereafter Cen~A, could dominate the observed anisotropy signal \citep{AbdulHalim:JCAP:2024}. 
In addition, the arrival directions of UHECRs $\gtrsim32~$EeV suggest a hotspot in the direction of Cen A 
\citep{Abreu:ApJ:2022,AbdulHalim:2023p}.

Jets from Active Galactic Nuclei (AGN) have long been considered as promising UHECR sources 
\citep[e.g.,][]{Hillas1984ARA&A,BlandfordPhST2000,Aharonian2002PhRvD}. 
Recent studies have shown that the UHECR spectrum and anisotropy could be related to their relativistic 
jets \citep[e.g.,][]{EichmannJCAP2018,MatthewsMNRAS2018,EichmannJCAP2022}.
However, the mechanisms responsible for accelerating particles to these extreme energies are still 
uncertain. 
In the case of the powerful, highly relativistic jets in Fanaroff-Riley (FR) type II radio galaxies, 
it has been suggested that UHECR acceleration could occur at their fast jet termination shocks  
\citep{RachenBiermann,Cerutti2023A&A,Huang2023MNRAS}, at sub-relativistic shocks in the jets' back-flows \citep{Matthews2019MNRAS,BellMNRAS2019} or at a sharp velocity transition on the jets' edge \citep{Ostrowski1998A&A,Rieger2004ApJ,Caprioli2015ApJL,Kimura2018,Mbarek2021,SeoApJ2023}.
In less powerful, mildly-relativistic FR~I type sources, UHECR acceleration by stochastic processes 
has been considered to occur within their large-scale lobes \citep{Hardcastle2009MNRAS,OSullivan09} 
or in gradual velocity-shearing jet flows \citep{Rieger2004ApJ,Rieger2007Ap&SS,Liu2017ApJ,Webb2018,Webb2019,
Rieger2019ApJL,Lemoine2019,Webb2020,Wang2021MNRAS,Rieger2022ApJ,Wang2023MNRAS}. Relativistic magneto-hydrodynamic (RMHD) simulations show that Kelvin–Helmholtz driven sheaths form 
at the interface between the jet spine and the cocoon, indicating that shear acceleration may in fact 
be a generic mechanism in mildly relativistic jets \citep{Wang2023MNRAS}.

The UHECR candidate source Cen A is the nearest FR~I radio galaxy at a distance of $D\approx3.8$~Mpc 
\citep{Harris2010PASA} and well-studied across the electromagnetic spectrum \citep[e.g.][]{Israel1998}. 
High-energy observations of its kpc-scale jet favor a synchrotron origin of the X-ray and an inverse 
Compton origin of the TeV emission \citep{H.E.S.S.Collaboration2020Natur}. These findings imply that 
efficient (in-situ) acceleration of electrons takes place along the length of its jet. Shear 
acceleration of particles provides a natural explanation for this \citep{Wang2021MNRAS}. Interestingly,
the observed jet speeds in Cen~A are rather mildly relativistic ($\beta\sim0.5-0.7$) \citep[e.g.,][]
{Hardcastle2003ApJ,Snios2019ApJ}, disfavoring acceleration processes that rely on large ($\Gamma>5$) 
% FR: would be inclined to keep this given the recent "claim" for superluminal motion in Cen A, 
% suggesting that \Gamma~3 may occur
bulk Lorentz factors. 
Similar constraints apply to many FR~II sources where radio observations indicate rather mildly 
relativistic bulk velocities for their kpc-scale jets \citep[e.g.][]
{WardleMNRAS1997,HardcastleMNRAS1999,ArshakianMNRAS2004,
MullinMNRAS2009,HardcastleNewAR2020}. Modeling of the kpc-scale jet emission in FR~II radio galaxies 
indeed shows that their radio to X-ray spectrum can be reproduced within a framework of electron shear 
acceleration with a jet bulk velocity $\beta\sim0.9$ \citep{Wang2021MNRAS,HeMNRAS2023}. These findings
suggest that efficient shear acceleration of electrons could occur in both FR I and FR II jets, and 
motivates a related analysis of its role for cosmic-ray energization.

In this Letter, we explore whether UHECRs can be accelerated in the large-scale jets of radio galaxies, 
taking Cen~A as a prime example. We perform dedicated RMHD simulations, tracking test particles using the 
PLUTO code \citep{Mignone2007ApJS,Mignone2018ApJ}. 
Our setup differs from previous studies that explored strongly magnetized flows or highly relativistic 
jets \citep[e.g.,][]{Mbarek2021,Medina2023}.  
We explicitly focus on input parameters for the magnetic field and velocity as inferred from earlier
analytical modelling of multi-wavelength observations of Cen~A, and employ high-resolution 
simulation to properly probe the growth of instabilities and gyro-motion of the test particles. 
In Section \ref{sec:setup}, we introduce our simulation setup. The results are shown in Section 
\ref{sec:results}.
The conclusion and implication on understanding the origin of UHECRs are provided in Section 
\ref{sec:Conclusion_discussion}.

\section{Simulation setup}\label{sec:setup}
In order to probe the acceleration process, RMHD simulations are performed, building on previous 
findings \citep{Wang2023MNRAS}, but using a significantly increased resolution. The simulation details 
are provided in Table \ref{tab:1}, summarizing the key parameters. 
The simulations are performed in a three-dimensional Cartesian geometry with the jet's bulk motion 
directed along the $y$-axis. The simulation volume is $2l_0\times l_0\times 2l_0$, where length $l_0$ 
and grid resolution $\Delta l$ are defined in Table \ref{tab:1}. 
The cylindrical jet is initialized with a uniform axial velocity $\beta_0$ in the spine with radius 
$\rj$, i.e., \(\bm{\beta} = \beta_0 \bm{\hat{y}}\) for $r<\rj$ and zero otherwise. Small transverse velocity 
perturbations are applied at the jet edge \citep{Rossi2008A&A}. 
Periodic boundary conditions are adopted in the $y$ direction, with outflow boundary conditions used 
in the $x$ and $z$. 
%The current jet power of Cen A is estimated to be $\sim10^{43}$~erg/s.

To mimic the kpc-scale jet of Cen A, we set $\beta_0=0.6$ \citep{Hardcastle2003ApJ,Snios2019ApJ} and 
label these simulations as `FR Ia, Ib, Ic'. Radio and X-ray observations \citep[e.g.,][]{Hardcastle2003ApJ,
Hardcastle2006MNRAS,Kraft2009ApJ} suggest that this large-scale jet has a radius of order $\rj=0.1$~kpc 
and a (projected) length of about $4.5$~kpc. On smaller scales, the inclination angle of the jet axis 
wrt the line of sight is constrained to $12^\circ-45^\circ$ \citep[e.g.,][]{Janssen2021}. This suggests 
a physical (de-projected) jet length $Z_{\rm CenA}\sim (6-22)$~kpc, and a minimum propagation time 
$t_{\rm jet, min}\approx Z_{\rm CenA}/\beta_0 c\sim (110-360) \rj/c \sim 10^5\,{\rm yrs}$. 

To enable comparison with FR II type jets, we also simulate a jet with $\beta_0=0.9$ \citep[e.g.][]
{WardleMNRAS1997,HardcastleMNRAS1999,ArshakianMNRAS2004,MullinMNRAS2009,HardcastleNewAR2020}, 
labelled `FR II'. For this scenario, the initial jet radius is $\rj=1$~kpc and is allowed to extend
to several hundreds of kpc in length \citep[e.g.][]{ZensusARA&A1997,WilsonApJ2001,HarrisARA&A2006,
JesterMNRAS2007,SchwartzIAUS2015,BlandfordARA&A2019,Wang2021MNRAS,HeMNRAS2023}, or equivalently a
jet propagation time of at least several hundred $\rj/c$. Particle trajectories are tracked up to 
$t=1000\rj/c$.

\begin{figure}
    \centering
    \includegraphics[width=0.49\textwidth]{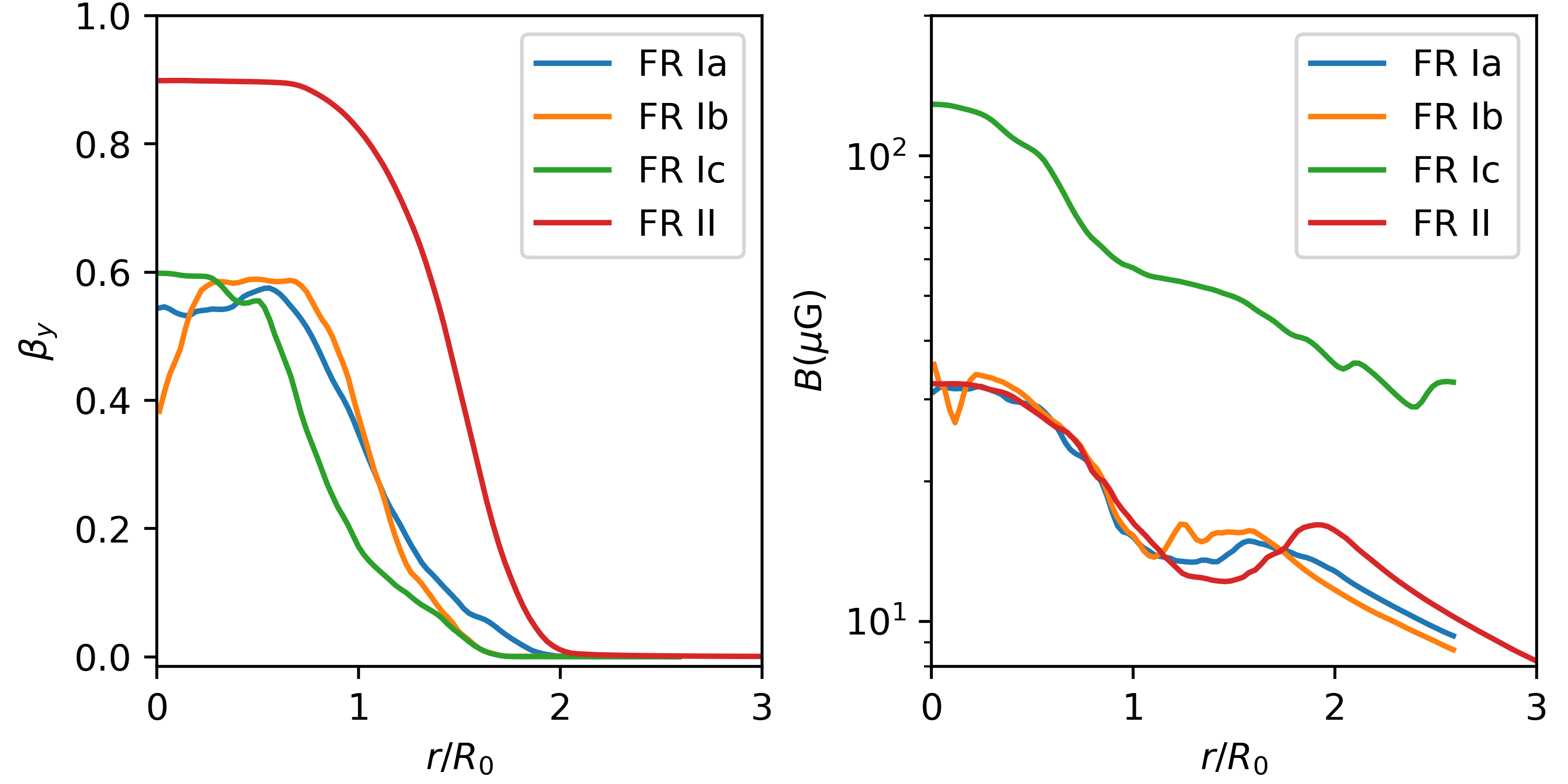}
    \caption{The velocity (left panel) and magnetic field (right panel) profiles in the radial 
    direction are averaged over the axial and azimuthal directions for the simulated jets at 
    the frozen time $t=\tf$.}
    \label{fig:profiles}
\end{figure}

We simulate jet propagation in a static cocoon. The jet and cocoon are initialized with pressure 
balance and a temperature $\Theta \equiv p_{\rm g}/n_0 m_{\rm p} c^2<1$ using a Taub-Mathews equation of state
\citep{Mathews1971ApJ,Mignone2005ApJS,Mignone2007ApJS}, where $p_{\rm g}$ is the gas pressure and $m_{\rm p}$ 
is the proton mass. 
The proper number density of the jet is $n_0= 10^{-6}\,{\rm cm}^{-3}$ and the cocoon density is 
$2n_0$. 
For the magnetic field we adopt a helical profile with $<B_{y}^2> = <B_{\phi}^2>$ for the axial and 
toroidal components respectively, and define the magnetization parameter $\sigma= \bar{B}_0^2/(8\pi
\rhoj c^2)$, where the average value $<f> \equiv \int_0^{\rj} 2\pi f r dr/\int_0^{\rj} 2\pi r dr$ 
assumes axial symmetry \citep{Wang2023MNRAS}. 
The initial mean magnetic field in the spine ($\bar{B}_0=\sqrt{<B_{y}^2+ B_{\phi}^2>}$) and the 
temperature on the jet axis ($\Theta_0$) are listed in Table \ref{tab:1}.
Multi-wavelength emission modelling of Cen A suggests a sheath magnetic field strength of 
$\sim 20\mu$G \citep{H.E.S.S.Collaboration2020Natur,Wang2021MNRAS}. We take this value for
guidance, but allow for some variation in view of the uncertainties in analytic models, and 
the fact that the observed jet radius of Cen~A grows with jet length \citep{Hardcastle2006MNRAS}.
We have studied two magnetizations ($\sigma=0.02,~0.2$) for our FR Ia-c simulation runs, where the 
\rm simulated sheaths have nonuniform magnetic fields in the range of $\sim10-60~\mu$G (see Figure 
\ref{fig:profiles}).

In our test-particle simulations, we inject energetic protons as test particles at $t=0$ to co-evolve 
with the RMHD simulations. 
The test particles are injected over the whole jet length within a radial zone $\Delta r_{\rm inj}$, 
with the same injection Lorentz factor $(\gamma_{\rm inj})$ as listed in Table \ref{tab:1}. 
In each cell within the injection region, we inject one particle with a random initial direction. 
The injection energy is chosen such that the particle Larmor radius is few grid cells in 
length, $\bar{r}_{\rm L, inj}\sim (2-4) \Delta l$, and thus energetic enough to sample the 
resolved structures formed in the sheath. 
In the case of Cen~A, seed energization could be facilitated by particles being (pre-)accelerated 
along the jet as indicated by multi-wavelength observations \citep{Hardcastle2006MNRAS,Kataoka2006,
H.E.S.S.Collaboration2020Natur}, 
or by external entrainment of cosmic ray particles \citep{Caprioli2015ApJL,Kimura2018}.
In total, we have $N_{\rm inj}=(0.9-4.4)\times10^7$ test particles in each 
simulation. %43582000 14522000 8742000
Different particle injection zones have been tested within $r_{\rm inj}\leq1.05\rj$ for the FR Ia
simulation run (see Appendix \ref{app:injection_radius}). 
The particle injected within $r_{\rm inj} \in [0.74,0.9]\rj$ are accelerated to slightly higher energies, but overall the difference in spectra is small. 
Thus we consider particles injected at $r_{\rm inj} \in [0.8,0.9] \rj$ for other simulation runs.

The RMHD simulation and test-particle simulation are advanced in time simultaneously until a time ($t= \tf$), which are listed in Table \ref{tab:1}.
Afterwards, the RMHD simulation is frozen while the test-particle simulation continues to run in the
static magnetic and electric fields. The frozen times are selected to be in the saturated KHI stage, 
where the turbulent kinetic energy of the jet stays almost constant \citep{Wang2023MNRAS}.
In the framework of Fermi II and shear acceleration, acceleration can occur along the whole jet length. 
In this manner, the choice of the frozen time reflects the acceleration of particles at various locations 
along the jet with different velocity and magnetic field profiles.

\begin{table}[h] %    \centering
    \begin{tabular}{ ccccccccc }
    \hline
Runs&$\beta_0$ &$\tf$  & $R_0$ & $l_0 $ & $\Delta l$  & $\bar{B}_0$  & $\Theta_0 $ %&  $L_{\rm K}$(erg/s) 
& $\gamma_{\rm inj}$ \\%&  $r_{\rm inj}(R_0)$
& & $(R_0/c)$ & (kpc) & $(R_0)$ & (pc) &  $(\mu{\rm G})$ &
\\\hline
FR Ia & 0.6 & $120 $ &0.1 & $2.6$ &0.5   & $27.4$ &  $0.05$  %& $1.4\times10^{43}$ 
& $5\times10^7$\\% &  
FR Ib & 0.6 & $60 $ &0.1& $2.6$ &0.5  & $27.4$ &  $0.05$ % & 
& $5\times10^7$ \\%&
FR Ic & {0.6} & {160} & {0.1} & {2.6} & {0.5} & {86.9} & {0.05}  &$9\times10^7$ \\

FR II & 0.9 & $60 $&1 & $3.0$ &6.0 & $27.4$ &  $0.09$  % & $7.9\times10^{45}$ 
& $3\times10^8$ \\%& 
%FR IIb & 0.99 & $100 $&1 & $4.0$ &7.8 & $27.4$ &  $0.09$  % & $7.9\times10^{45}$ & $4\times10^8$ \\%& 
\hline
    \end{tabular}
    \caption{Key parameters used in the simulations: initial jet-spine velocity ($\beta_0$) and radius ($\rj$), MHD freeze-time ($\tf$), simulation box size ($l_0$), grid size 
    ($\Delta l$), initial mean magnetic field $\bar{B}_0$, initial jet temperature parameter $\Theta_0$, and initial Lorentz factor of injected protons ($\gamma_{\rm inj}$).}
    \label{tab:1}
\end{table}

\begin{figure}
    \centering
    \includegraphics[width=0.49\textwidth]{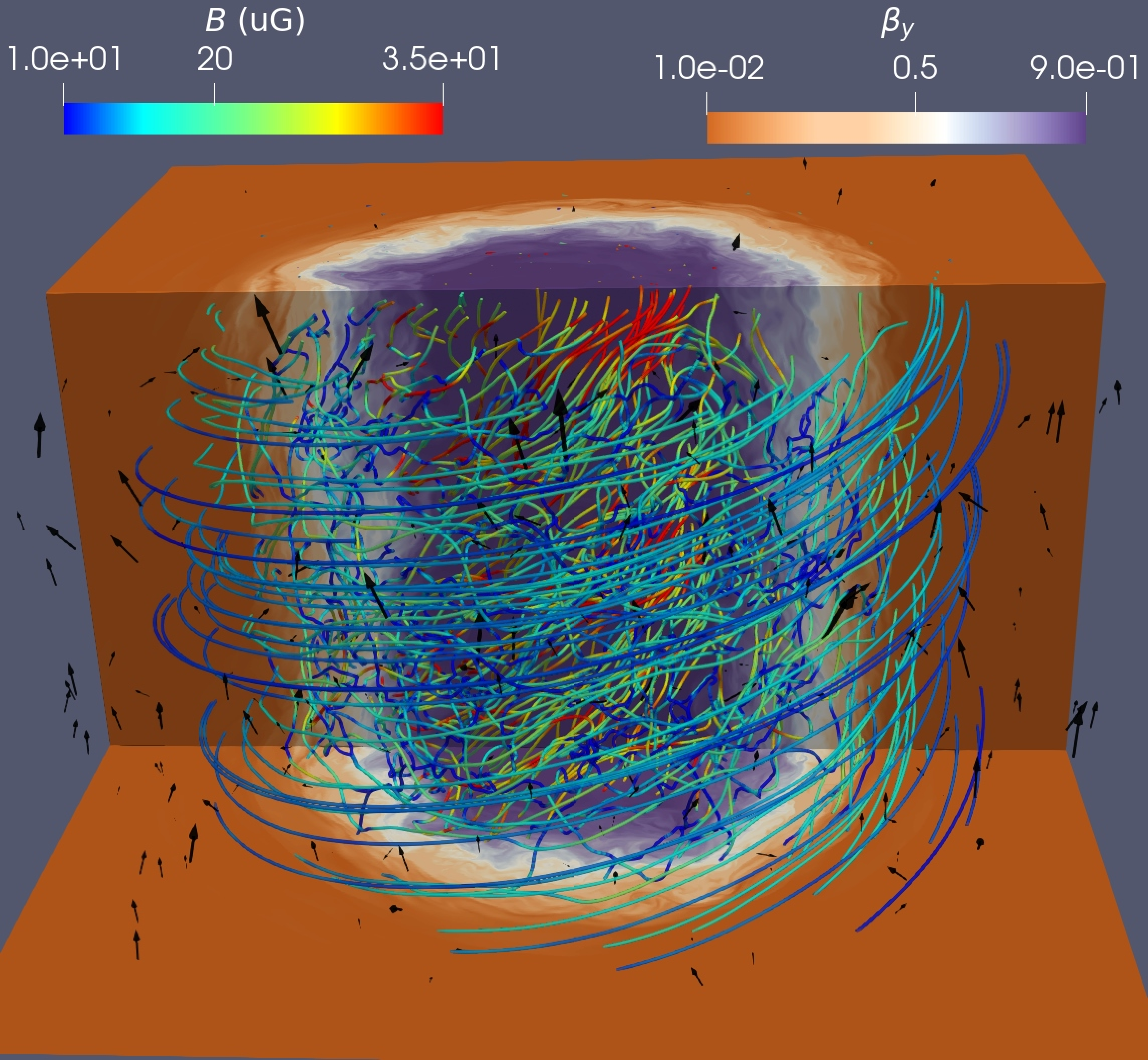}
    \caption{An example of simulation run FR II at $t=\tf$. The box size is 
    $6\rj\times3\rj\times6\rj$. 
    The distribution of the axial velocity 
    $\beta_y$ is shown in order to illustrate the spine-sheath jet and cocoon structure.
    Magnetic field lines are shown with color coding of their magnitude. 
    The black arrows represent test particles with larger arrow sizes for 
    higher-energy particles.}
    \label{fig:mhdpic}
\end{figure}

\section{Simulation results}\label{sec:results}
Figure \ref{fig:mhdpic} shows an example of the magnetic field structure and velocity profile 
in a simulation of an FR II type jet. While the behaviour of the jet-sheath 
dynamics and the developed turbulence have been studied previously \citep{Wang2023MNRAS}, they 
are now probed over a substantially wider range.
The KHI operates at the interface between the spine and the cocoon, and facilitates formation 
of a turbulent spine-sheath structure. 
At the saturated KHI stage, the velocity turbulence spectra are consistent with 
Kolmogorov scaling over nearly two decades (see Appendix \ref{app:turb}), 
while the magnetic turbulence exhibits some flattening towards small wavenumbers. 
To maximise the dynamic range particles are injected at low energies with Larmor radii close to the grid scale. While numerical diffusion results in deviation from 
Kolmogorov-scaling at the grid-scale, as shown in Figure \ref{fig:Turb}, the results are found to be 
consistent when injecting at higher energies, as discussed in Appendix \ref{app:injection_energy}.
Therefore, we fix the injection energy for other simulations as shown in Table \ref{tab:1}.
Averaging over the axial and azimuthal directions, we show examples of the velocity and magnetic 
field profiles along the radial direction of the spine-sheath jet at $t=\tf$ of the simulation 
runs in Figure \ref{fig:profiles}. 
Since the jet radius changes over time, we hereinafter re-define the spine-sheath jet radius as 
the region with $\beta_y\geq0.01$, i.e., 
$R_{\rm j}= r(\langle \beta_y\rangle =10^{-2})$.
In the sheath, the velocity drops smoothly, while the magnetic field displays a pile-up at the 
edge of the jet in the low-magnetization simulation runs.

\begin{figure}[h]
    \centering
    \includegraphics[width=0.49\textwidth]{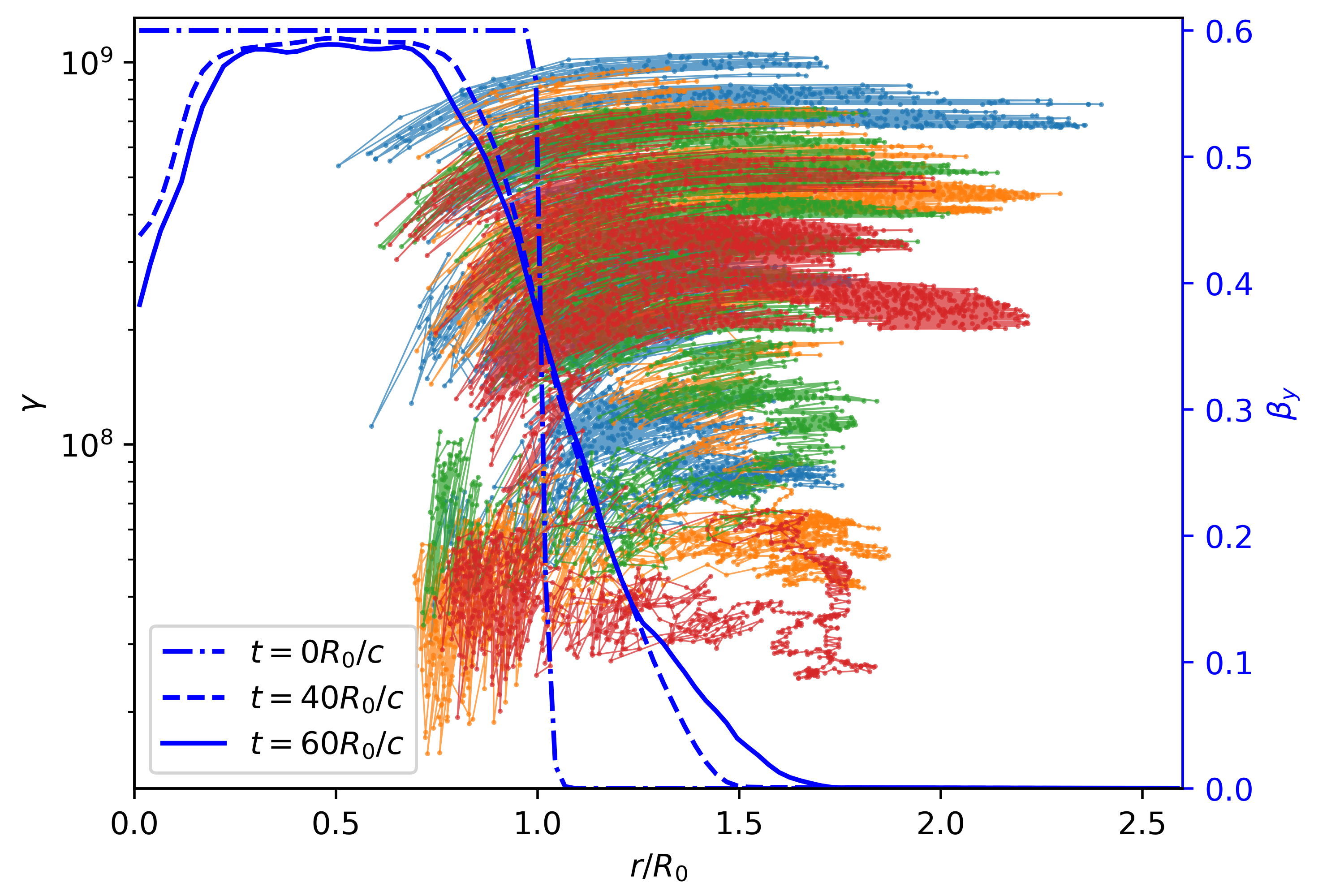}
    \caption{Exemplary trajectories of four test particles are shown with different colors for 
    the FR Ib simulation run.
    The trajectories are projected in the radial direction of the jet. 
    The left $y$-axis is the Lorentz factor ($\gamma$) of test particles.
    The blue lines are the velocity profiles of the jet at different simulation times, with the 
    values shown at the right $y$-axis. }
    \label{fig:v6t60-trajec}
\end{figure}

In such a shearing turbulent jet sheath, particles are expected to undergo stochastic Fermi II 
acceleration as well as (gradual) shear acceleration. The latter can also be viewed as 
a stochastic Fermi-type acceleration process, where charged particles gain energy by scattering 
off the turbulence embedded in a large-scale sheared flow profile \citep[e.g.,][]{Rieger2019Galax,Lemoine2019}. 
Figure \ref{fig:v6t60-trajec} provides an illustration of these effects. 
As particles repeatedly traverse the turbulent jet-sheath, they gain energy. To study the potential 
for UHECR acceleration in such scenarios, we focus on the spectrum and the maximum energy of the 
simulated particles below. For the analysis of the particle spectra, we take only particles inside 
the jet (at $r\leq R_{\rm j}$) into account.

\begin{figure}[h]
    \centering
    \includegraphics[width=0.7\textwidth]{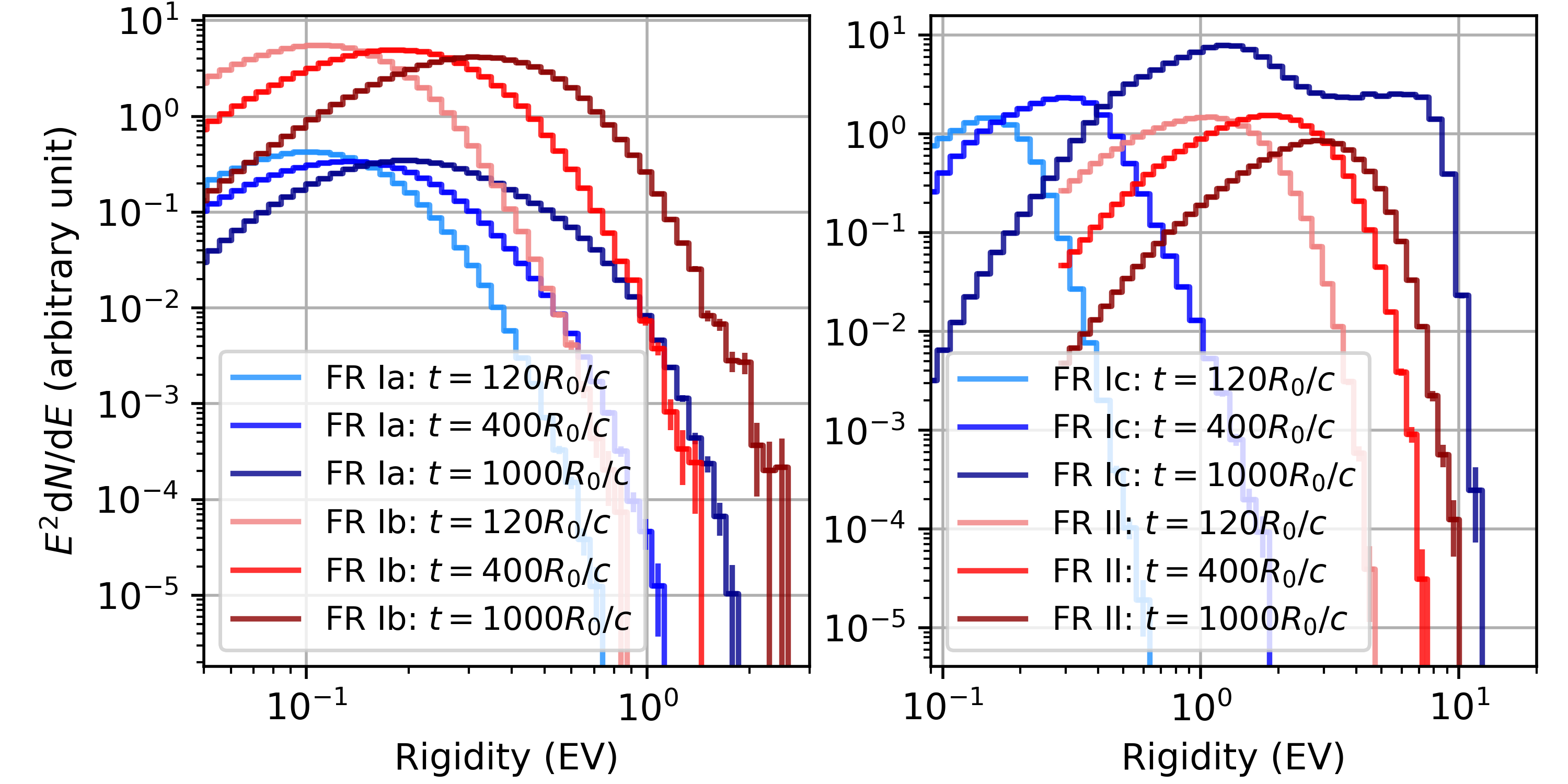}
    \caption{The normalized particle spectra of test particles as obtained at different 
    simulation times $t=120 \rj/c$, $t=400 \rj/c$ and $t=1000 \rj/c$ for the simulation 
    runs FR Ia, Ib (left panel), Ic and FR II (right panel). 
    The normalization factors are chosen differently 
    for the purpose of visualization. 
    To avoid overlapping lines, we only show the particle 
    spectrum above the injection energy.}
    \label{fig:v6-v9-CR-SED-Time}
\end{figure}

In Figure \ref{fig:v6-v9-CR-SED-Time} we show the time evolution of the particle spectra for 
the FR I and FR II simulation runs.
For the low-magnetization FR I jets, we employ two simulation runs at a different MHD frozen 
time: $\tf=60\rj/c$ (FR Ib) and $\tf=120\rj/c$ (FR Ia). 
Although protons are injected mono-energetically in the simulation frame, the particle distributions 
broaden over time, spanning an energy range more than a factor of $10$ above the injection energy. 
This is typical for stochastic acceleration processes. 
From the FR Ia/b simulation runs, we found that the particle spectra do not change significantly 
with different choices of frozen time. 
Thus we perform only one simulation for an FR I jet with a higher magnetization (FR Ic), and one 
simulation for FR II jets. 
For both, the FR I and FR II cases, the spectral peak energy and the maximum particle energy 
increase systematically due to acceleration, with roughly the same spectral shape at $t\geq120\rj/c$. 
These results demonstrate that efficient particle acceleration by stochastic processes is 
operative in mildly relativistic jets with speeds $\beta\sim0.6-0.9$.

\begin{figure}
    \centering
    \includegraphics[width=0.5\linewidth]{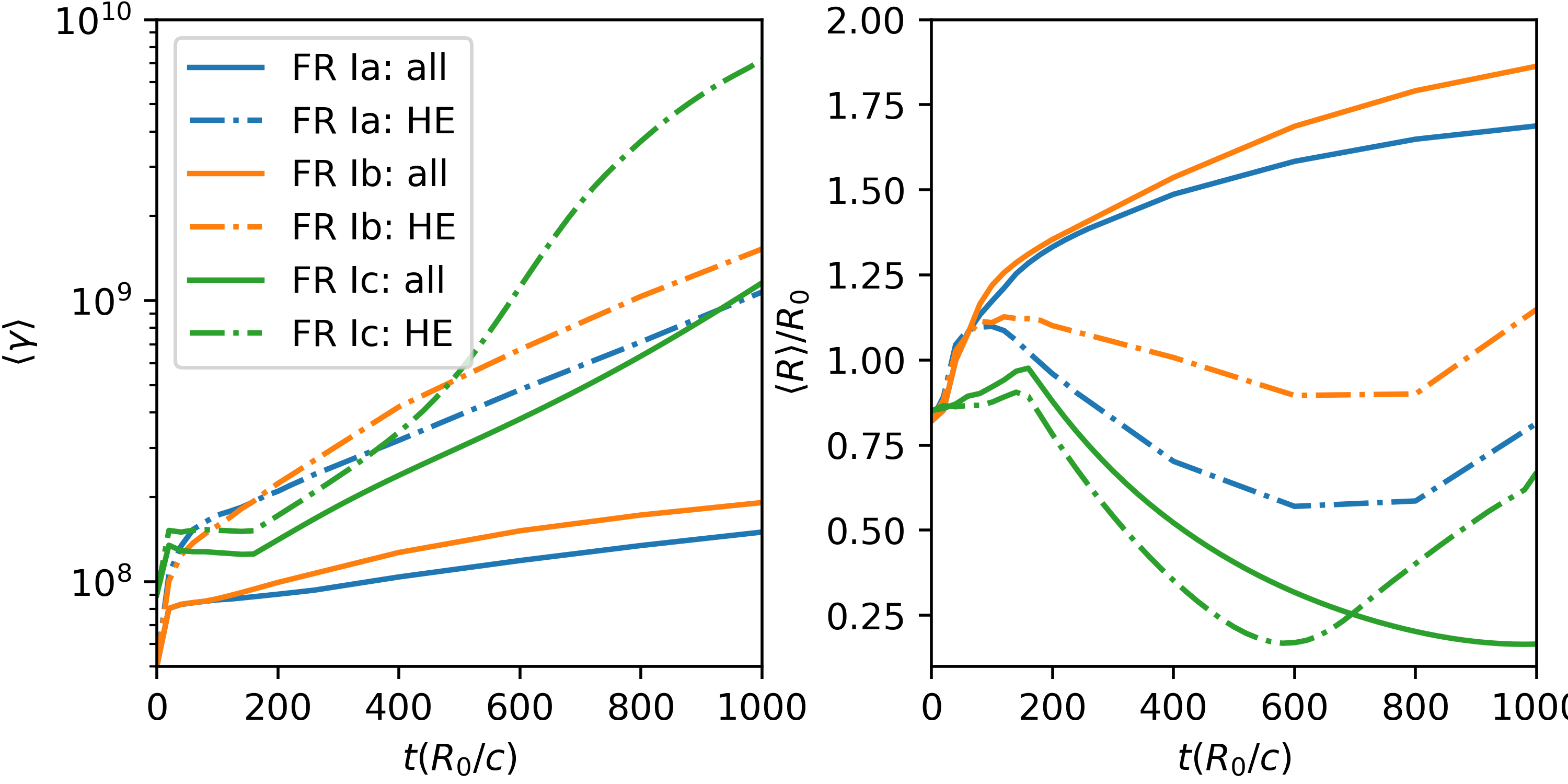}
    \caption{The time evolution of mean energy (left panel) and mean radial position 
    (right panel) for the FR I simulations. 
    The right panel has the same figure legend as the left panel.
    The solid lines, labeled with `all' are for the all-particle sample. 
    The dashed-dotted lines, labeled with `HE' are for the high-energy component.}
    \label{fig:E-R-time}
\end{figure}

For the FR Ic simulation, an additional high-energy (HE) component becomes apparent at 
later time, as shown in Figure \ref{fig:v6-v9-CR-SED-Time}. 
To explore the difference between the highest-energy particles and all the particles 
for the FR I simulations, we investigate the time evolution of the acceleration rate 
and acceleration location. 
We select particles with energy $\gamma>0.5\gamma_{\rm max}$ for the respective HE 
component, where $\gamma_{\rm max}$ is the highest-energy particle in each simulation.
We subsequently obtain the time evolution of the mean energy ($\langle \gamma\rangle$) 
and mean radial position ($\langle R\rangle$) for both, the all-particle sample and the 
HE sample. The results are shown in Figure \ref{fig:E-R-time}. As can be seen, the 
evolutionary trends for the FR Ia/b simulations are quite similar, while the FR Ic 
simulation behaves differently. At later time $t\gtrsim 400 R_0/c$, the HE component 
of the FR Ic simulation reveals an increased acceleration compared to the other 
simulations.
We attribute this visual difference to the higher magnetization employed in our FR Ic
simulation that contributes to a separation of stochastic (Fermi II) and shear 
particle acceleration processes.
The mean radial positions show that the all-particle samples systematically moves to 
the outer/inner edge of the sheath, where in FR Ia/b simulations the trend is to move 
outwards with $\langle R\rangle/R_0\gtrsim1.7-1.9$ and to move inwards with $\langle R
\rangle/R_0 \lesssim0.2$ in the FR Ic simulation. However, in general the HE components 
in these simulations tend to stay in the sheath, experiencing both stochastic and shear 
acceleration. The HE component of the FR Ic simulation reveals increased acceleration 
compared to the all-particle sample, primarily occurring when those HE particles move 
from the spine to the inner sheath, while the all-particle sample continues moving 
inward. The higher velocity (gradient) and/or the higher magnetic field in the inner 
sheath may thus contribute to the observed difference in energization. 
We plan to investigate this in more detail with further simulations in a future paper.

\begin{figure}
    \centering
    \includegraphics[width=0.49\textwidth]{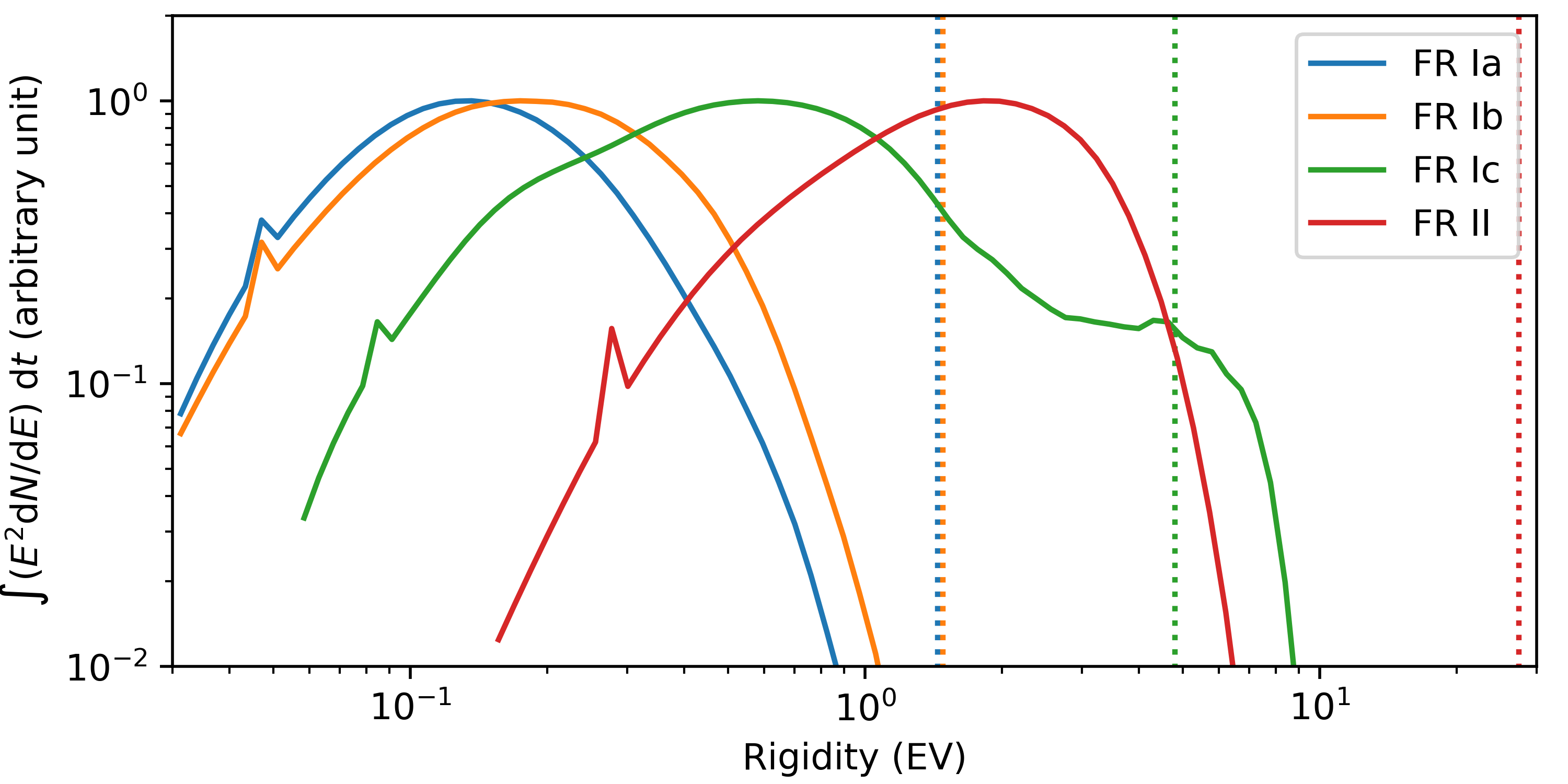}
    \caption{The spectra are integrated over time to mimic continuous particle 
    injections for our simulation runs. The vertical lines denotes the Hillas 
    reference limit for the jet.
    The spectra are normalized such that the first data points are equal to unity.
    The spikes in the spectra are an artefact of the mono-energetic injection 
    (see $\gamma_{\rm inj}$ in Table \ref{tab:1}).}
    \label{fig:sed-integrated}
\end{figure}

As particles are injected impulsively at $t=0$ in our simulations, we integrate the spectrum over 
time to mimic continuous injection. To quantify the acceleration capability, we define a Hillas 
limit \citep{Hillas1984ARA&A} for particles in our simulated jets, $E_{\rm max}=q \beta B 
R_{\rm j}$, where $R_{\rm j}$ is the jet radius at the frozen time $t=\tf$. The integrated particle
spectra for all simulation runs are shown in Figure \ref{fig:sed-integrated} along with dotted lines 
for $E_{\rm max}$. 
It is seen that for different jet magnetizations and velocities, the particle energy spectra peak 
at around $0.1 E_{\rm max}$. Below the peak, the time-integrated spectra are as hard as ${\rm d}
N/{\rm d} E\propto E^{-1}$, which are compatible with efficient stochastic-shear acceleration and 
ineffective escape. A more complex situation concerning particle escape could arise in the case of
significant pre-existing magnetic turbulence in the cocoon.
In the case of shear acceleration, the acceleration efficiency depends on the velocity profile of 
the jet with a sharper velocity-shearing flow (larger flow gradients), leading to more efficient 
acceleration \citep[e.g.,][]{Webb2018,Rieger2019ApJL}. As shown in Figure \ref{fig:profiles} the 
FR Ib simulation run exhibits a steeper velocity gradient in the sheath, which could account for 
the difference in the spectra compared to that of the FR Ia simulation run. 
For the FR Ic simulation, individual particles do reach beyond the noted Hillas limit. 
Note, however, if we take the jet diameter instead of the radius as the system size, those particles would 
be within the limit.

\section{Conclusions}\label{sec:Conclusion_discussion}

In this letter, we have explored the possibility for a source such as Cen A to 
accelerate particles to multi-EeV energies via stochastic Fermi-type processes.
Cen A is particularly interesting as the UHECR candidate source that could 
provide the dominant contribution to the observed anisotropy signal 
\citep[e.g.,][]{AbdulHalim:JCAP:2024}. %RomeroAPh1996
Based on observations of its current large-scale jet, the available time for 
the acceleration of particles is at least $t_{\rm jet, min} \sim (110-360) \rj/c$ 
as estimated above. 
For a jet speed of $\beta_0=0.6$ and a magnetization of ${\sigma=0.02-0.2}$ (FR I 
simulations), our simulations show that the resultant proton energy spectra 
($E^2dN/dE$) peak at ${\sim0.2-0.7}$~EeV, and extend beyond $1$~EeV. The spectral 
peak approximately satisfy $E_{\rm peak} \approx 0.1  E_{\rm max}$ for 
simulations with different magnetization and velocity parameters. 
As suggested in the literature \citep[e.g.,][]{MatthewsMNRAS2018},
Cen A may have exhibited enhanced activity in the past accounting for its 
giant radio lobe structure. Our results indicate that in the presence of larger 
jet velocities and/or higher jet magnetizations, possibly associated with such 
a phase, the peak in the proton energy distribution could be shifted to higher 
energies. Futher studies on the multiwavelength jet properties of Cen A will be 
important to quantify this in more detail.

Analysis of the PAO observations shows that the amplitude of the UHECR dipole 
anisotropy increases from $1.7\%$ in $4-8$~EeV to $17\%$ at $\geq 32$~EeV 
\citep{PierreAugerCollaboration2017Sci,Aab2018ApJ,AbdulHalim:2023p}. 
The mass composition at those energies is likely a mix of H, He, and CNO-type 
elements \citep{AbdulHalim:2023Yd,Coleman:APh:2023}.
To account for this by Cen A, FR Ic type simulations are favored, where the 
rigidity spectrum extends well beyond several EV. 
Heavy elements could in principle enter the jet through mass entrainment by, 
e.g., stellar material inside the jet \citep{Wykes:MNRAS:2015,Bosch-Ramon:A&A:2023}, 
or from the medium surrounding the jet shearing layer \citep{Perucho:Galax:2019}. 
Additionally, lower-energy cosmic rays produced within the host galaxy 
may be picked up by the jet, contributing to variations in the mass 
composition \citep{Caprioli2015ApJL,Kimura2018}. It seems likely, however, that 
to properly account for the inferred super-solar UHECR abundance, some further 
enhancement of heavy elements at the injection scale is needed. To which extent 
this might be facilitated by an additional pre-acceleration mechanism remains 
to be investigated.
While the accelerated particles in our simulations have a hard spectrum, 
UHECRs escaping from the jet-cocoon system could reveal an even harder 
spectrum. This seems in general consistent with the suggested hard UHECR
spectrum \citep[e.g.,][]{Aab:JCAP:2017,Ehlert:PhRvD:2023}.

While offering important insights, the amplitude of the UHECR anisotropy is 
$\lesssim10\%$, indicating that there is a dominant isotropic component. 
The recent combined analysis of the UHECR spectrum and composition data 
favours a rigidity cutoff at a few~EV \citep{Aab:JCAP:2017,Aab2020PhRvL,
AbdulHalim2023JCAP}. It has been suggested that the isotropic UHECR component 
could originate from more powerful FR II type radio galaxies 
\citep[e.g.,][]{EichmannJCAP2018,MatthewsMNRAS2018,EichmannJCAP2022}. 
Our results provide an illustration that for mildly relativistic FR II type 
conditions, (rigidity-spectrum) peak energies around $\sim2$~EV for $t=
(120-400)\rj/c$ can be reached. 
It seems thus likely that UHECRs at $100~$EV could represent heavy elements 
from powerful FR II jets exhibiting higher magnetization and/or larger jet 
radii.

\begin{acknowledgments}
We are grateful to the referee for suggestions.
J.S.W. thanks T. Bell and M. Lemoine for discussions.
J.S.W. acknowledges support by the Alexander von Humboldt 
Foundation and F.M.R by the German Science Foundation 
(DFG RI 1187/8-1). 
%The simulations were performed on the HPC system Raven at the Max Planck Computing and Data Facility.
\end{acknowledgments}

\vspace{5mm}
\software{
PLUTO \citep{Mignone2007ApJS,Mignone2018ApJ}; NumPy \citep{harris2020array}; Matplotlib \citep{Matplotlib}, ParaView \citep{ParaView}
} 

\facility{The HPC system Raven from the Max Planck Computing and Data Facility (MPCDF)}

\appendix
\restartappendixnumbering

\section{The effect of injection radius}\label{app:injection_radius}
\begin{figure}
    \centering
    \includegraphics[width=0.5\linewidth]{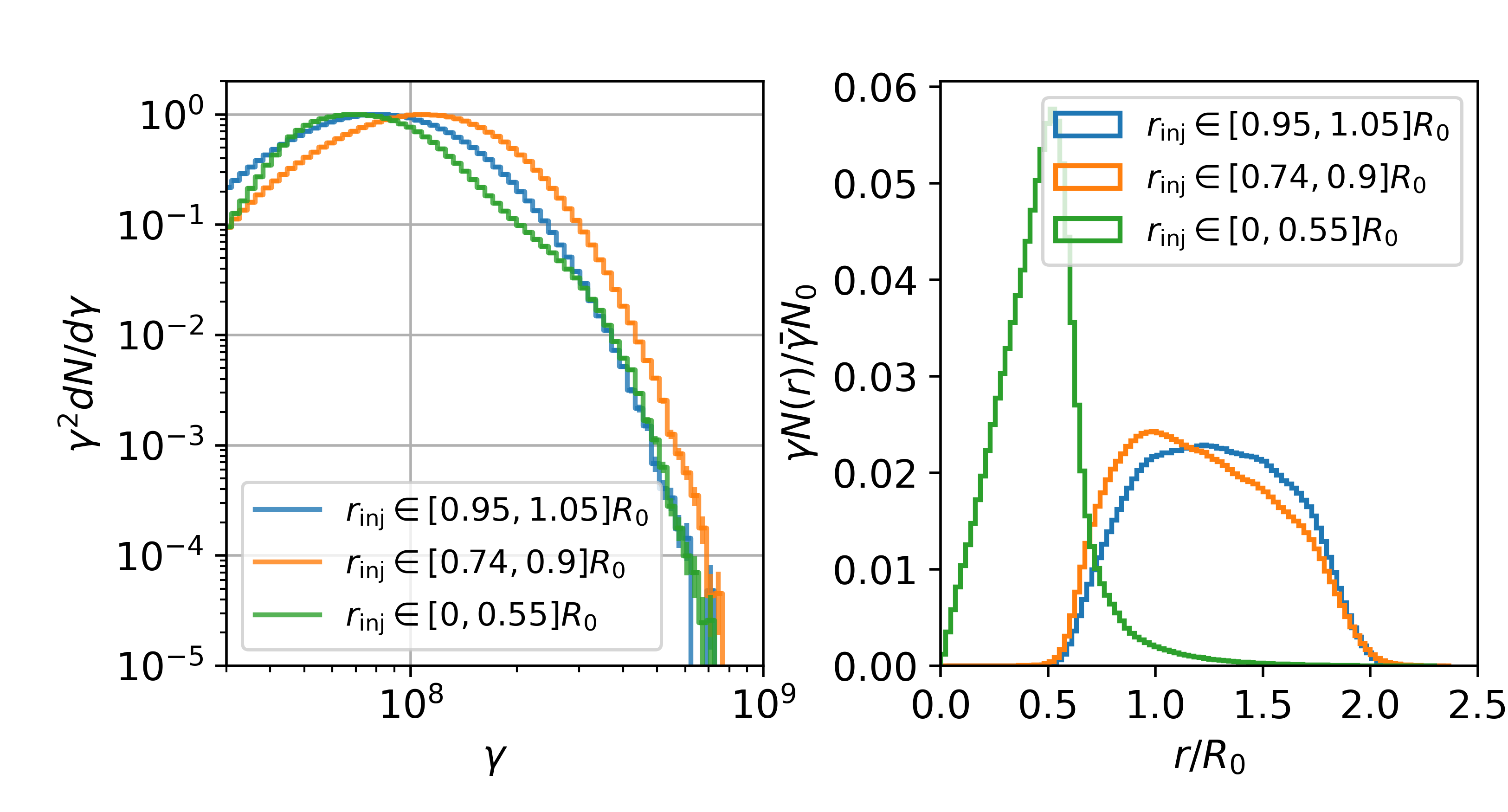}
    \caption{The spectral energy distributions (left panel) and spatial distributions 
    (right panel) are shown for particle injected at different radius at the frozen time 
    $\tf=120 \rj/c$.
    The particle energy spectra are normalized to unity at the peak value, while the spatial distribution are normalized so that the sum of distribution equals to unity.}
    \label{fig:Compare-injection-R}
\end{figure}

For the FR Ia simulation run, we have injected particles in the region $r_{\rm inj}\in[0,0.55]\rj,~
[0.74,0.9]\rj,~[0.95,1.05]\rj$. 
The test particles co-evolve with the RMHD simulation to the frozen time.
At the frozen time, the particle spectral and the spatial distribution are shown in 
Figure \ref{fig:Compare-injection-R}.
It can be seen that particles spread out spatially to a wide radius range, especially for 
those particles injected close to the jet boundary, while the difference in the energy 
spectra is overall small.
Particles injected at $r_{\rm inj}\in [0.74,0.9]\rj$ have a peak in the inner sheath $r\sim\rj$, 
and are accelerated to slightly higher energies compared to the other two cases.
This is in general consistent with stochastic-shear acceleration, as the inner sheath has 
a slightly larger velocity gradient.

\section{Turbulence spectra}\label{app:turb}
\begin{figure}
    \centering
    \includegraphics[width=0.5\linewidth]{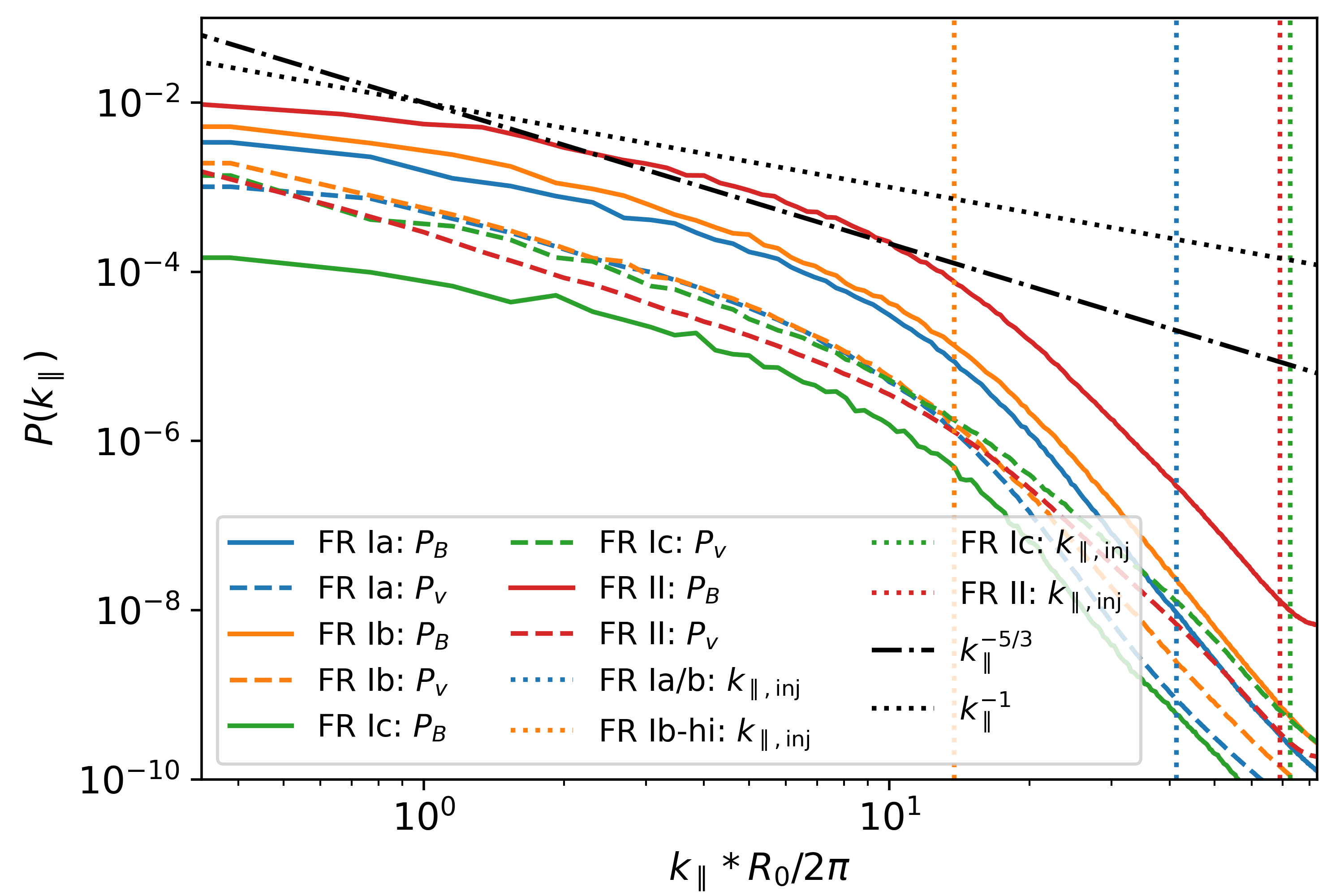}
    \caption{The turbulence spectra for the velocity ($P_v$) and magnetic field ($P_B$) 
    are shown for the different simulation runs.
    The vertical dotted lines corresponds to the average Larmor radii of the injected energy for different simulations (see Sections \ref{sec:results} and Appendix \ref{app:injection_energy} for details). }
    \label{fig:Turb}
\end{figure}

Following \cite{Wang2023MNRAS}, we perform a fast Fourier transform of the velocity and 
magnetic fields along the jet axis with wavenumber $k_{\parallel}$ at the saturated KHI stage ($t=\tf$). 
The minimum and maximum wavenumber are determined by the simulation box-size and grid size, 
respectively.
The results are shown in Figure \ref{fig:Turb}. 
It can be seen that the power-law component of the power spectra generally extends over 
approximately two decades within wavenumbers $ k_{\parallel} R_0/2\pi\lesssim 10-20$.
The turbulent velocity spectra are compatible with a Kolmogorov-type turbulence, while 
the magnetic turbulence reveals some flattening towards small wavenumber, approaching a
Bohm scaling ($k_{\parallel}^{-1}$).

\section{The effect of injection energy}\label{app:injection_energy}

\begin{figure}
    \centering
    \includegraphics[width=0.5\linewidth]{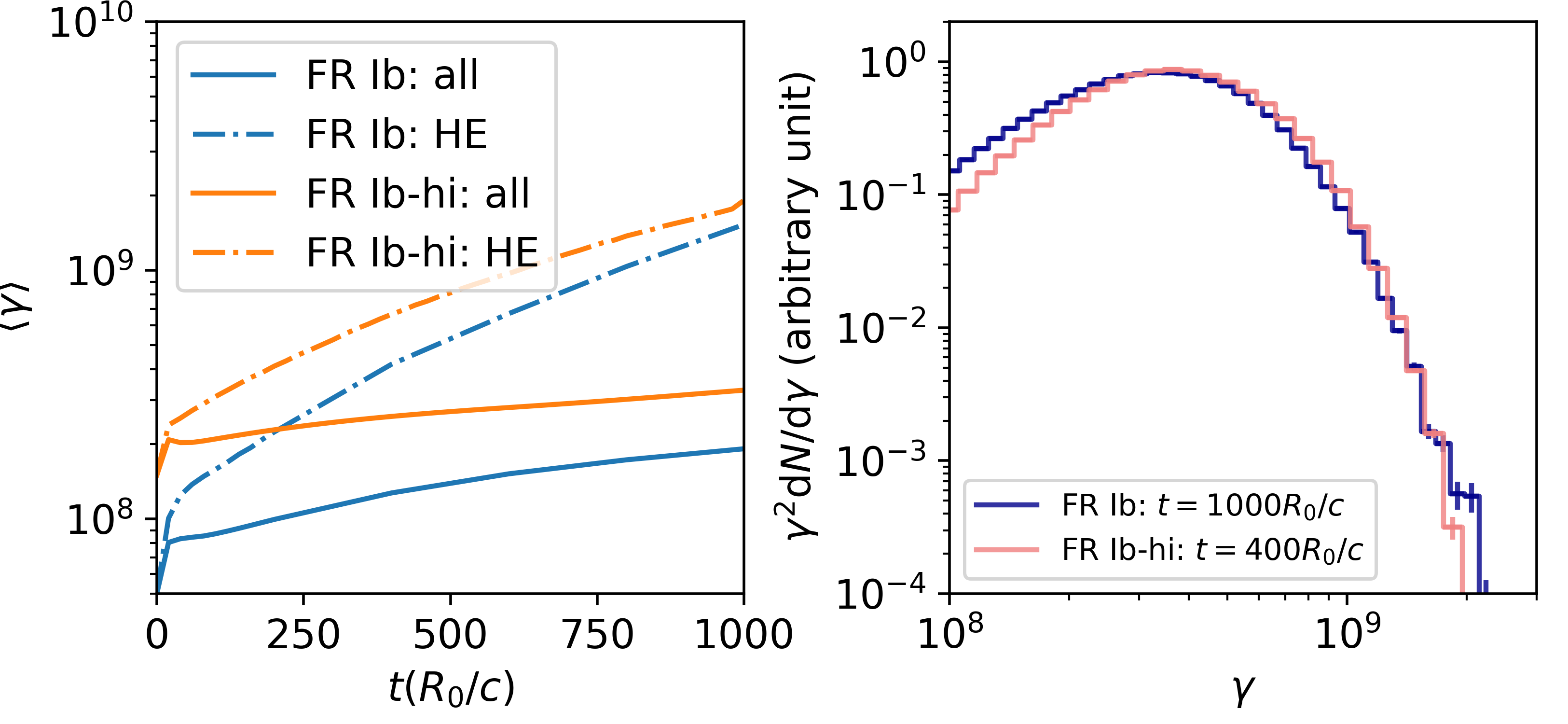}
    \caption{
    The time evolution of the mean particle energy (left panel) and the spectra (right panel) of the FR Ib and FR Ib-hi simulations. 
    The definition of the all-particle sample and HE component in the left panel are the same as in Figure \ref{fig:E-R-time}. 
    In the right hand panel, the normalization and particle energy range are selected for visualization purposes, similar to Figure \ref{fig:v6-v9-CR-SED-Time}. 
    }
    \label{fig:compare-inj-gamma}
\end{figure}

As shown in Figure \ref{fig:Turb}, the Larmor radii of the injected particles in the simulations resonate with wavenumbers $k_{\parallel,\rm inj} R_0/2\pi \approx 40-70$, which lies beyond Kolmogorov-scaling range. This is unavoidable if we inject close to the grid scale.
To determine if this changes any of our conclusions, we thus conducted an additional simulation, FR Ib-hi, based on the initial setup of the FR Ib simulation, only with a higher injection Lorentz factor $\gamma_{\rm inj}=1.5\times10^8$, corresponding to an average resonant wavenumber $k_{\parallel,\rm inj} R_0/2\pi \approx 14$. 
We present the time evolution of the mean particle energy and spectra in Figure \ref{fig:compare-inj-gamma}. 
The mean particle energy in FR Ib simulation reaches $\langle \gamma\rangle\approx 1.5\times10^8$ at $t\approx600 R_0/c$. 
Therefore, we compare the results between FR Ib simulation at $t\gtrsim600 R_0/c$ and FR Ib-hi simulation at $t\lesssim 400  R_0/c$.
Our results show that the mean acceleration rates ($d\log \langle \gamma\rangle/dt$) for both the all-particle sample and HE component and the spectra for FR Ib simulation at $t=1000 R_0/c$ and FR Ib-hi simulation at $t=400 R_0/c$ are comparable.
This indicates that variations in injection energy do not significantly affect the results. 
The reason could be that the particle spectra in the FR Ib simulation peak above a Lorentz factor of $1.5\times10^8$ at $t\gtrsim120 R_0/c$ (Figure \ref{fig:v6-v9-CR-SED-Time}), and the average particle energy reaches this scale in a time $t\gtrsim600 R_0/c$.
These particles can experience the Kolmogorov-type turbulence with $k_{\parallel} 
R_0/2\pi \lesssim 14$ for a long duration of time,
such that the result resembles that of the FR Ib-hi simulation with an injection energy 
$\gamma_{\rm inj}=1.5\times10^8$.

\bibliography{ref}{}
\bibliographystyle{aasjournal}

\end{CJK*}
\end{document}